\documentclass[preprint]{JHEP3}
  
\def\to{\rightarrow}
\def\be{\begin{equation}}
\def\ee{\end{equation}}
\def\bea{\begin{eqnarray}}
\def\eea{\end{eqnarray}}

\def\nn{{\nonumber}}
\def\sF{{F \hspace{-7pt}/}}
\def\spartial{{\partial \hspace{-6pt}/}}
\def\eda{{\eta^\dagger}}

\title{AdS$_3$ Solutions of IIB Supergravity from D3-branes}
 \author{Nakwoo Kim \\ 
 Department of Physics and Research Institute of Basic Science, \\ 
 Kyung Hee University, Seoul 130-701, South Korea \\
 E-mail: \email{nkim@khu.ac.kr}}
 \abstract{
 We consider pure D3-brane configurations of IIB string theory which lead to supersymmetric
 solutions containing an AdS$_3$ factor. They can provide new examples of AdS$_3$/CFT$_2$ examples
 on D3-branes whose worldvolume is partially compactified. When the internal 7 dimensional space
 is non-compact, they can be identified as supersymmetric fluctuations of higher dimensional 
 AdS solutions and are in general dual to 1/8-BPS operators thereof. We find that supersymmetry requires the
 7 dimensional space take the form of a $U(1)$  fibration over a 
 6 dimensional Kahler manifold.}
 
\begin{document}
 \section{Introduction}
 Due to the celebrated duality relation between string theory in anti-de-Sitter (AdS) backgrounds and conformal field theories on the boundary\cite{Maldacena:1997re}, 
it is nowadays of great interest to find
new AdS solutions to string/M-theory and study their field theory duals.
Several different avenues have been taken to accomplish this goal. 
For instance, when one replaces the spheres of the maximally
supersymmetric examples by squashed spheres new AdS/CFT duality
pairs can be obtained \cite{Acharya:1998db}. 
When the squashed sphere is a toric  
Sasaki-Einstein manifold, one can identify the dual gauge theory and 
nontrivial predictions of the duality relation can be shown to match 
the field theory results \cite{Martelli:2004wu}. 
In particular, the field theory computation of Weyl anomaly using the
a-maximization method \cite{Intriligator:2003jj} is shown to agree with the volume of toric
Sasaki-Einstein manifolds through Z-minimization developed 
in \cite{Martelli:2005tp}.
Another approach which proved to be 
very fruitfull is to use the lower-dimensional gauged supergravity theories 
as the springboard, i.e. first to find nontrivial AdS vacua in low dimensions 
and then uplift the solution to 10 or 11 dimensions.  They correspond to 
fixed points of the renormalization group equation confronted when the
maximally supersymmetric system is perturbed by a relevant operator.
See, for instance Ref. \cite{Pilch:2000fu}.

When we restrict ourselves to supersymmetric solutions we can
take a more systematic approach to study
the generic forms of such vacua as in \cite{Gauntlett:2002sc,Gauntlett:2002nw}. 
Instead of trying to solve the
Killing spinor equation employing a specific ansatz, one can derive the 
general form of the solution by making use of the information given by the Killing
spinor equations. One constructs various spinor bilinears from the 
Killing spinor and the algebraic and differential conditions derived from
the Killing spinor equation help to restrict the local form of the 
metric. For lower dimensional systems with less supersymmetry
it is sometimes possible to classify all supersymmetric solutions. Several
important class of new solutions have been found using this method. 
Supersymmetric black ring solutions and Sasaki-Einstein manifolds $Y^{p,q}$
\cite{Gauntlett:2004zh, Gauntlett:2004yd} 
are among them. A valuable insight into the gauge/gravity duality
has been obtained from the analysis of IIB string solutions containing
two factors of three-spheres with unbroken supersymmetry \cite{Lin:2004nb}. 
They are dual to the 1/2-BPS
operators of $N=4$ super Yang-Mills theory and 
it was shown that a two-dimensional slice of
the supergravity solution can be identified as fluctuations of free fermion phase space. 

In this work we study IIB vacua with a AdS$_3$ factor and establish
how supersymmetry restricts the local form of the metric and field
fluxes. In M-theory, a comprehensive analysis for supesymmetric
AdS$_3$ solutions has been undertaken in \cite{Martelli:2003ki}.
To simplify the analysis we consider pure D3-brane configurations,
i.e. the nontrivial fields of IIB supergravity are only Ramond-Ramond
5-forms and metric. We find that the transverse 7-dimensional space
takes the form of a $U(1)$ fibration over a complex 3-dimensional
Kahler space. If the whole 7-dimensional space is compact we obtain
a new AdS$_3$ vacua. Otherwise the solution should describe a
BPS operator of higher dimensional conformal field theory, most likely
4-dimensional. In the next section we present our setup in detail, and 
present how the supersymmetry provides relations between various
differential forms in the internal space. We also illustrate how 
the well-known AdS solutions can be reproduced when we choose
appropriate 6 dimensional Kahler manifolds. We conclude with
comments and suggestions for future works.

 \section{Ansatz}\label{ansatz}
It is the goal of this work to study supersymmetric IIB solutions with a
AdS$_3$ factor from D3-branes.
Among the various fields of IIB supergravity we allow to turn on the metric
and Ramond-Ramond five-forms only. We plan to derive how the existence of
a nontrivial Killing spinor solution restricts the local form of the metric
and the five-form field strength.
Having fixed a 3 dimensional part
of the metric, we get an effective system which is 7 dimensional. 

We introduce
a scalar field $A$ and a two-form field strength $F$ as follows.
\bea
ds^2 &=& e^{2A} (AdS_3) + g_{ab} dx^a dx^b , \quad a,b=1,2,\ldots 7. 
\label{metric}
\\
F^{(5)} &=& (1 + *) {\rm Vol}_{AdS_3} \wedge F .
\eea
The dilatino variation vanishes trivially and we only need to 
consider the gravitino variation equation,
\be
\nabla_M \epsilon +
\frac{i}{480} \Gamma^{M_1\ldots M_5} F^{(5)}_{M_1\ldots M_5} \Gamma_M 
\epsilon = 0 .
\ee

It is required to introduce a specific basis for the gamma matrices which
respect the dimensional decomposition we consider here. 
\bea
\Gamma_{\mu} &=& \sigma_1 \otimes \gamma_\mu \otimes 1 , 
\quad \mu = 0,1,2  . 
\\
\Gamma_a  &=& \sigma_2 \otimes 1 \otimes \gamma_a , 
\quad a = 3, \ldots 9 .  
\eea
In this basis the 10 dimensional chirality projection implies $\sigma_3\epsilon
=+\epsilon$. On AdS$_3$ the Killing spinor satisfies the following 
property.
\be
\nabla_\mu \epsilon = \frac{a}{2} \gamma_\mu \epsilon,
\quad a=\pm 1 . 
\ee
Now we can rephrase the Killing spinor equation in terms of a 7 dimensional
Dirac spinor $\eta$.
We obtain the following set of equations,
\bea
\nabla_a \eta - \frac{e^{-3A}}{4} \sF \gamma_a \eta &=& 0 , 
\label{k1}
\\
\Big(  \spartial A + \frac{e^{-3A}}{2} \sF -ia e^{-A} \Big) \eta &=& 0 . 
\label{k2}
\eea
In 7 dimensions with Euclidean signature 
it is possible to define
Majorana spinors but from the Killing equations above it is obvious 
we need to consider a Dirac spinor. Since $\eta$ has 8 components
generic Killing spinor solutions should preserve 1/8 supersymmetry. 
We might understand this statement as the supersymmetry of D3-branes
wrapping a Kahler two-cycle in a Calabi-Yau four-fold which consists of
the 2 tangential and 6 transverse directions of the D3-brane world-volume. 
%%%%%%%%%%%%%%%%%%%
\section{Spinor Bilinears}
%%%%%%%%%%%%%%%%%%%%
One can construct differential forms of various ranks 
defined on the 7 dimensional space as spinor bilinears.
\bea
C &=& \eda\eta ,\\
K_a &=& \eda\gamma_a\eta, \\
Y_{ab} &=& i \eda\gamma_{ab} \eta,\\
Z_{abc} &=& i \eda\gamma_{abc} \eta,\\
W_{abcd} &=&  \eda\gamma_{abcd} \eta,\\
X_{abcde} &=&  \eda\gamma_{abcde} \eta,\\
P_{abcdef} &=&  i\eda\gamma_{abcdef} \eta .
\eea
One can also consider complex conjugate spinors to construct differential
forms which are complex valued. Because of the antisymmetry of gamma
matrices only 3- and 4-forms are non-vanishing \footnote{
In our convention the 7 dimensional gamma matrices are all
antisymmetric.}, we define
\bea
\Omega_{abc} &=& \eta^T \gamma_{abc} \eta , 
\\
\Psi_{abcd} &=& \eta^T \gamma_{abcd} \eta . 
\eea
Now one can make use of the 7 dimensional Killing equations 
to derive algebraic and differential relations between the 
spinor bilinears. For instance,
\bea
\nabla_a (\eda\eta) &=& \nabla_a \eda \eta + \eda \nabla \eta
\nn \\
&=& - e^{-3A} F_{ab} \eda \gamma^b \eta
\nn \\
&=& \partial_a A \eda \eta ,
\label{sc}
\eea
which implies $\eda\eta=e^A$.

Proceeding in the same way one finds that $K$ in fact defines a Killing
vector, i.e. $\nabla_{(a} K_{b)} = 0$. We choose a local coordinate patch such 
that $K=\partial_\psi$ and write the 7 dimensional metric as
follows,
\bea
ds^2 &=&
g_{ab} dx^a dx^b \quad\quad  a,b=1,\ldots 7 . 
\nn\\
&=& 
 e^{2\phi} (d\psi + B)^2 + g_{ij} dx^i dx^j , 
\quad i,j = 1,\ldots , 6 .
\label{7d}
\eea
$\phi$ is a 6-dimensional scalar which is given as the
norm of $K$ through $K^2 = e^{2\phi}$  
and $B$ is a one-form in 6 dimensions. 

One can in fact see that $\phi=A$. 
From the algebraic Killing equation Eq.(\ref{k2})
it follows $\eta^T\eta=0$. 
If we introduce a pair of Majorana spinors to 
write
\be
\eta = \frac{1}{\sqrt{2}} ( \eta_1 + i \eta_2 ),
\nn
\ee
it follows that $\eta_1^T\eta_1=\eta^T_2\eta_2$ and $\eta_1^T \eta_2 = 0 $.
We also have $\eda\gamma^i\eta=0$, which implies 
$\eta_1^T \gamma^i \eta_2=0$.
Because of the orthogonality of gamma matrices 
and that they are 8 dimensional, it is obvious that 
$\eta, \gamma^i\eta, \gamma^\psi\eta$ span the whole spinor space.
Since $\eta_1$ is orthogonal to $\eta_2$ and $\gamma^i\eta_2$, one concludes
that $\eta_1 \propto \gamma_\psi \eta_2$. Since $\eta_i$
are real spinors of the same magnitude, we conclude 
$\eta_1=\pm i \gamma_{\hat{\psi}}\eta_2$,
which implies the Dirac spinor $\eta$ is chiral on 
6 dimensional space defined by $x^i$ in Eq.(\ref{7d})
\footnote{
Here we use the notation where the hatted indices denote orthonormal
frame, i.e. $\{ \gamma_{\hat{a}} , \gamma_{\hat{b}} \} = 2 \delta_{\hat{a}\hat{b}}$.}.
Then 
$K^2 = e^{2\phi} =
(\eda\gamma^\psi\eta)(\eda\gamma_\psi\eta) = (\eda\eta)^2=e^{2A}$.
The chiral spinor $\eta$ thus defines a $SU(3)$ structure on the 6 dimensional
base space, $Y$ defines an almost complex structure and $\Omega$
is a $(3,0)$-form. 

Above procedure of exploiting Killing equations can be repeated for 
other differential forms. One first uses the differential relation to 
compute the exterior derivative, and the result is simplified by making
use of the algebraic equation. We summarize
the result as follows,
\bea
d(e^{2A}K) &=& -4 F - 2a e^A Y ,
\label{ex1}
\\
d(e^A Y) &=& 0 ,
\label{ex2}
\\
d(e^{4A} Z) &=& 4a e^{3A} W - 4 e^A Y\wedge F ,
\\
d(e^{3A} W) &=& 0 ,
\\
d(e^{2A} X) &=& 2 a e^A P ,
\\
dP &=& 0 .
\eea
Now from the chirality condition of $\eta$ in 6 dimensions all
the higher forms can be written
as a wedge product of $K,Y$ and only the first two equations in 
the above are independent. One also derives 
\be
d(e^{2A}\Omega) = 2ai K \wedge \Omega .
\label{hol}
\ee

Now one can easily see that for the rescaled metric 
$\bar{g}_{ij}=e^{2A}g_{ij}$, $J=- e^AY$ and
$\omega = e^{2A}e^{2ai\psi}\Omega$ provide the canonical two-form 
and the $(3,0)$-form of the almost complex structure
defined by $Y$. In particular, using Fierz identity, one can check that
\bea
{\rm Vol}_6 &=& \frac{1}{6} J\wedge J\wedge J 
\nn
\\
&=&
 \frac{i}{8}
  \omega \wedge \bar{\omega} . 
\eea
$dJ =0$ and $d\omega = 2aiB\wedge \omega$
together then imply that the complex structure is integrable and $2 a dB$ is the Ricci-form
of the Kahler manifold.  

There is another equation derived from Eq.(\ref{k2}) when
we multiply $\eta^T$, 
which relates the 6 dimensional curvature scalar with $A$. 
We do not find any other independent conditions from the Killing equations
Eq.(\ref{k1},\ref{k2}), so
we here summarize the equations which guarantee the supersymmetry
of the configuration,
\bea
F &=& \bar{F} + K \wedge e^{2A} dA , 
\\
a e^{4A} {\cal R} &=& -8 \bar{F} + 4a J , 
\\
R &=& 8 e^{-4A} . 
\label{cur}
\eea
where $\bar{F}$ is the two-form field restricted to 6 dimensional space
and $J^{ij} = \bar{g}^{ik}\bar{g}^{jl}J_{kl}$. ${\cal R}$ is the
Ricci form, and $R$ is Ricci scalar. It is clear
that once the Kahler space is fixed the above equations 
can determine $A,F$, thus the entire 10 dimensional solution. 

It is an established fact that a supersymmetric configuration satisfies
the classical field equations provided the form-field equations of motion 
and the Bianchi identities are satisfied \cite{Gauntlett:2002fz}. The Bianchi
identity $dF=0$ is a consequence of supersymmetry as can be easily
seen from Eq.(\ref{ex1}) and Eq.(\ref{ex2}). The form-field
equation of motion $d(e^{-3A} *F)=0$ 
can be checked most easily using the second line of 
Eq.(\ref{sc}). As we take another covariant derivative, after some algebra
one can show it leads to 
\be
\Box R - \frac{1}{2} R^2 + R_{ij} R^{ij} 
= 0 
\label{dim}
\ee
where the norms are taken with respect to the rescaled metric $\bar{g}$.

Let us summarize. We have shown that, if we restrict ourselves to pure
D3-brane backgrounds, any supersymmetric solution of IIB supergravity with 
an $AdS_3$ factor can be always written,
\bea
ds^2 &=& e^{2A} ds^2(AdS_3) +e^{2A} (d\psi+B)^2 
+ e^{-2A} ds^2_{{\rm Kahler}} 
\\
F^{(5)} &=& (1+*) {\rm Vol}_{AdS_3} \wedge \left( \frac{a}{2} J
- \frac{1}{4} d( e^{4A} (d\psi + B)  \right) 
\eea
In principle, one can construct new $AdS_3$ solutions starting with 
a 6 dimensional Kahler space satisfying Eq.(\ref{dim}), then 
using Eq.(\ref{cur}) and $dB=2{\cal R}$. Eq.(\ref{dim}) can be 
rewritten as a 
4th-order partial differential equation for the Kahler potential.
Instead of trying to solve Eq.(\ref{dim}) directly, 
in the
remainder of this article we illustrate how well-known solutions
can be rephrased in terms of our result. 

%%%%%%%%%%%%%%%%%%%%%%%%
\section{Examples}
%%%%%%%%%%%%%%%%%%%%%%%%%%
We now construct explicit solutions from the
equations presented in the last section and in particular show how the 
well-known AdS solutions can 
be rephrased in our general framework. 

As the simplest case we choose the 6 dimensional Kahler basis to be
Einstein or products of Kahler-Einstein spaces, i.e. 
$A={\rm const}$. Eq.(\ref{dim}) then becomes a simple algebraic
relation involving the dimensionalities of Kahler-Einstein manifolds,  
\be
R^2=2R_{ij}R^{ij} 
\label{concase}
. 
\ee
It is easy to see that the only possibility is $S^2\times T^4$ if we exclude
the use of hyperbolic spaces. This case corresponds to the most well-known 
 example of IIB supergravity with a AdS$_3$ 
factor, i.e. AdS$_3 \times S_3 \times T^4$ which has 1/2 unbroken 
supersymmetry. When we set $A=0$ the
radius of $S^2$ is 1/2, and if we introduce $\theta,\phi$ as the
coordinates on $S^2$ with the standard metric, 
the 7 dimensional metric is written as
\bea
ds^2 &=& \frac{1}{4} \Big(
(d\psi + \cos\theta d\phi)^2 + d\theta^2 + \sin^2\theta d\phi^2
\Big)
% \nn\\  && 
+ (T^4 \mbox{-part}) .
\eea
The 10 dimensional solution is interpreted as the near-horizon limit
of two intersecting D3-branes on a string. 

As the second example we consider the type of solutions AdS$_5 \times SE_5$ 
where $SE_5$ is a 5 dimensional Sasaki-Einstein space. The Ramon-Ramond
five-form is given as the sum of AdS$_5$ and $SE_5$ volume form.
 These solutions
correspond to the near horizon limit of D3-branes put on a singular 
point of Calabi-Yau space. It is well known that a Sasaki-Einstein 
space can be always written as a Hopf-fibration over a Kahler-Einstein
space, i.e.
\be
ds^2 = (d\alpha + \sigma/3)^2 + ds^2_{KE} , 
\ee
with $d\sigma/6$ the Kahler form of the Kahler-Einstein base.
The constant norm Killing vector $\partial_\alpha$ is called
Reeb vector.

The simplest examples of Kahler-Einstein space are given from
complex projective spaces CP$^n$. In 4 dimensions we have
two obvious choices, CP$^2$ and CP$^1 \times$CP$^1$. 
The former gives rise to $S^5$, the latter $T^{1,1}$ respectively.
The metric cone of $T^{1,1}$ is the conifold and the dual gauge 
theory living on the D-branes put on conifold singularity is
understood in detail \cite{Klebanov:1998hh}.

Until recently $T^{1,1}$ has been the only 5 dimensional 
Sasaki-Einstein manifold whose metric is known explicitly.
A couple of new, infinite class of Sasaki-Einstein manifolds 
were discovered recently \cite{Gauntlett:2004yd,Cvetic:2005vk}. 
They all turned out to be toric, and the dual conformal 
field theories given as quiver gauge theories have been 
identified. 

It is possible to reconstruct these solutions from our equations. It
is just a matter of choosing the right 6 dimensional Kahler space.
In order to determine the right Kahler space we try to 
rewrite AdS$_5 \times SE_5$ solutions. 
\bea
ds^2 &=&
ds^2(\mbox{AdS}_5) + ds^2(\mbox{SE}_5)
\nn\\
&=&
\cosh^2 \rho \, ds^2 (\mbox{AdS}_3) + d\rho^2 + \sinh^2 \rho d\phi^2
%\nn\\ && 
 +\left( d\alpha + \frac{\sigma}{3} \right)^2 + ds^2 (\mbox{KE}_4)
\nn\\
&=&
\cosh^2 \rho \, ds^2 (\mbox{AdS}_3) +
\cosh^2 \rho \left( d\phi+ \frac{d\tilde{\alpha} + \sigma/3}{\cosh^2 \rho} 
\right)^2
\nn\\
&& +\frac{1}{\cosh^2\rho} \left(
\cosh^2\rho (d\rho^2 + ds^2 (\mbox{KE}_4)) 
%\nn\\ &&
+
\sinh^2 \rho (d\tilde{\alpha}+\frac{\sigma}{3})^2
\right)
,
\eea
where we set $\tilde{\alpha}=\alpha-\phi$. 
The Kahler form of the 6 dimensional
base space is written as follows, 
\be
J = \cosh\rho \sinh\rho d\rho \wedge (d\tilde{\alpha}+\sigma/3 )
 + \sinh^2\rho J_{KE} . 
\ee
 $J_{KE}$ is the Kahler form of the 4 dimensional Kahler-Einstein
manifold which gives rise to the Sasaki-Einstein manifold.
One can check $\cosh^2\rho$ and 
$2d\Big(\frac{d\tilde{\alpha}+\sigma/3}{\cosh^2\rho} \Big)$
corrrectly give the scalar curvature and the Ricci-form of the 6
dimensional Kahler space.

The next example is the 1/2-BPS solutions of IIB supergravity
obtained in \cite{Lin:2004nb}. Solutions with $SO(4)\times SO(4)$ symmetry,
i.e. having two factors of three-sphere are studied and argued to be dual
to generic 1/2-BPS operators of $N=4$ super Yang-Mills theory. Our
result can be easily translated into the case of IIB solutions with one 
$S^3$ instead of AdS$_3$ through double Wick rotation. The spacelike
Killing vector becomes a timelike Killing vector fibred again over a 
6 dimensional Kahler manifold. The solution takes the following form,
\bea
ds^2 &=& -h^{-2} (dt+V_i dx^i)^2 + h^2 ( dy^2 + dx^i dx^i) + y e^G d\Omega^2_3
+ y e^{-G} d\tilde{\Omega}^2_3 ,
\nn\\
h^{-2} &=& 2y \cosh G ,
\nn\\
y \partial_y V_i &=& \epsilon_{ij} \partial_j z, 
\quad
y (\partial_i V_j - \partial_j V_i) = \epsilon_{ij} \partial_y z ,
\nn\\
z &=& \frac{1}{2} \tanh G ,
\nn\\
F &=& dB_t \wedge (dt +V) + B_t dV + d\hat{B} ,
\nn\\
\tilde{F} &=& d\tilde{B}_t \wedge (dt +V) + \tilde{B}_t dV + d\hat{\tilde{B}} ,
\nn\\
B_t &=& -\frac{1}{4} y^2 e^{2G},\quad\quad \tilde{B}_t = -\frac{1}{4} y^2 e^{-2G},
\nn\\
d\hat{B} &=& -\frac{1}{4} y^3 *_3 d(\frac{z+1/2}{y^2}),\quad\quad 
d\hat{\tilde{B}} = -\frac{1}{4} y^3 *_3 d ( \frac{z-1/2}{y^2} ) , 
\label{llm1}
\eea
where $F,\tilde{F}$ are defined from the dimensional reduction of
Ramond-Ramond 5-form through
\be
F_{(5)} = F_{\mu\nu} dx^\mu \wedge dx^\nu \wedge d\Omega_3 
 + \tilde{F}_{\mu\nu} dx^\mu \wedge dx^\nu \wedge d\tilde{\Omega}_3 
 .  
 \ee
 The full solution is determined by a single function $z(x^1,x^2,y)$, which 
 satisfies the following equation. 
 \be
 \label{llm2}
 \partial_i\partial_i z + y \partial_y ( \frac{\partial_y z}{y} ) = 0 . 
 \ee
 
 In order to give a regular solution it is argued in \cite{Lin:2004nb}
 that $z=\pm \frac{1}{2}$ on 
 the $y=0$ plane. It turns out that once the the shape of
 the filled region defined by $z=\frac{1}{2}$ on the $(x^1,x^2)$-plane is
 specified the full 10 dimensional solution is determined. Then the
 filled region is interpreted as the fermi see of the (fermionized)
 Yang-Mills eigenvalues. Since our result applies to any supersymmetric
 solutions, any solutions satisfying Eqs.(\ref{llm1}),(\ref{llm2})
   can be written based on a 6 dimensional Kahler 
 manifold.

 We first write the metric of $\tilde{S}^3$ in terms of left-invariant
 forms
 \be
 d\tilde{\Omega}^2_3 = \frac{1}{4} ( \sigma^2_1 +\sigma^2_2 + \sigma^2_3 ),
 \ee 
 and introduce polar coordinates for the 2 dimensional space $(x^1,x^2)$
 as 
 \be 
 dx^i dx^i = dr^2 + r^2 d\phi^2 . 
 \ee
 In order to see the hidden Kahler structure it turns out useful to mix
the two Killing vectors $\partial_t$ and $\sigma_3$. When we rewrite
 $\sigma_3 \to \sigma_3 + 2 dt$, the metric can be written as 
 \bea
 ds^2 &=&
  y e^G d\Omega^2_3 -  y e^G ( dt + \frac{V}{h^2ye^G}  - \frac{\sigma_3}{2e^{2G}} )^2
  + \frac{1}{ye^G} 
  \Big[
  h^2 y e^G ( dr^2 + r^2 d\phi^2 ) + 
  \nn\\
  && 
  \frac{y}{4h^2 e^G} ( \sigma_3 - 2 V )^2 
  + h^2 y e^G dy^2 
  + \frac{y^2}{4} (\sigma^2_1 + \sigma^2_2 )  
  \Big]
 \eea
 The Kahler form is given as
  \be
 J = -(z+\frac{1}{2}) r dr \wedge d\phi + \frac{y^2}{4} \sigma_1 \wedge \sigma_2
 + \frac{y}{2} dy \wedge ( \sigma_3 - 2V ) . 
 \ee 
 One can easily check it is indeed closed, and compute the Ricci-form to 
 show that it is given as $d(\frac{2}{2z+1}V-e^{-2G}\sigma_3)$, 
 and Eq.(\ref{dim}) is satisfied.

\section{Discussion}
In this work we analyzed the D3-brane configurations which lead to AdS$_3$ 
solutions. The full 10 dimensional solution is determined once a 6 dimensional
Kahler space is chosen which satisfies a certain condition for the curvature,
given in Eq.(\ref{dim}). With appropriate Kahler manifolds we can reconstruct
known IIB solutions which contains an AdS$_3$ factor. Let us emphasize
that, since the higher dimensional AdS spaces can be written as a bundle 
over a lower dimensional AdS, we can describe all higher dimensional AdS
spaces from 5-forms in terms of our result. 

It would be very interesting to construct new AdS$_3$ solutions using
our result and
analyze their gauge theory duals. One can use hyperbolic
Kahler manifolds as part of the 6 dimensional Kahler base manifold to find
more solutions to Eq.(\ref{concase}). This is far from surprising, and in 
fact very reminiscent of the general
results of wrapped brane solutions reported in \cite{wrapped}, where
numerous AdS solutions have been obtained through wrapping branes on
hyperbolic supercycles. 

It is straightforward but very interesting
to consider the Wick rotated case of our solutions:
instead of assuming an AdS$_3$ factor one considers $S_3$. 
Supersymmetry and the $SO(4)$ isometry imply that in the dual picture
these solutions can describe pure scalar operators of $N=4, D=4$
super Yang-Mills which are in general
1/8-BPS. They can also describe 1/2-BPS operators of certain $N=1, D=4$
superconformal field theories. From the analysis in Ref.\cite{Lin:2004nb}
we know that part of the string theory spacetime can be mapped to
the phase space of the Yang-Mills eigenvalue dynamics. According
to our result supersymmetry requires that the phase space of
eigenvalue dynamics for generic BPS states should have not
only symplectic, but Kahler structure. It will be very exciting if 
one can find a prescription to extract the information on the dual
gauge theory and the fluctuation thereof, from the 6 dimensional
Kahler geometry. The geometric constraint Eq.(\ref{dim}) 
would tell us how the eigenvalue distribution is transcribed into
general relativity. 
The relevant matrix model description of the 
eigenvalue dynamics for 
$N=4, D=4$ Yang-Mills theory has been
discussed in \cite{Berenstein:2005aa}. We hope to be able to 
address this issue further in the near future.

\section*{Acknowledgements}
The preliminary version of this article has been reported at various places. 
We would like to thank the organisers of PASCOS-05 at Gyeongju, YITP 
workshop "String Theory and Quantum Field Theory" at the Yukawa Institute, 
and the Korean Physical Society meeting at Chonbuk university. 
It is also a great pleasure to thank  Jerome Gauntlett, 
Tetsuji Kimura, Dario Martelli, Liat Maoz, Asad Naqvi, Daniel Waldram, 
Jung-Tay Yee and Sang-Heon Yi
for discussions and valuable comments.  
This work was supported by the Science Research Center Program of
the Korea Science and Engineering Foundation 
(KOSEF) through the Center for Quantum Spacetime (CQUeST) of Sogang
University with grant number R11-2005-021, and 
by the Basic Research Program of KOSEF with grant No. R01-2004-000-10651-0.

\end{document}